\documentclass{article}

\usepackage[english]{babel}
\usepackage[utf8]{inputenc}
\usepackage[T1]{fontenc}

\usepackage[a4paper,top=2.5cm,bottom=3cm,left=2.5cm,right=2.5cm,marginparwidth=1.75cm]{geometry}

\usepackage{amsmath}
\usepackage[round]{natbib}
\usepackage{graphicx}
\usepackage[colorlinks=true, allcolors=blue]{hyperref}
\usepackage[detect-weight,retain-explicit-plus]{siunitx} %

\usepackage[yyyymmdd]{datetime}

\title{Ambisonic Encoding of Signals From\\ Equatorial Microphone Arrays}
\author{Jens Ahrens}
\date{Technical note\footnote{Find more technical notes at \url{http://www.ta.chalmers.se/education/texts-on-acoustics/}}~~v.~1\\[1ex] 
Chalmers University of Technology\\[1ex] 
\texttt{jens.ahrens@chalmers.se}}

\renewcommand{\d}{\mathrm{d}}
\renewcommand{\i}{\mathrm{i}}

\begin{document}
\maketitle

\begin{abstract}
The equatorial microphone array presented in (Ahrens et al., 2021) computes a spherical harmonic (SH) representation of a sound field based on pressure sensors along the equator of a rigid spherical baffle. The original formulation uses complex-valued SH basis functions. This is inconvenient if the SH representation of the captured sound field is intended to be stored in time domain by means of real-valued audio signals as it is common in the spatial audio format of ambisonics. The present document summarizes the modifications that need to be applied to the mathematical formulation from (Ahrens et al., 2021) to produce an ambisonic representation of the captured sound field that is compatible with the established ambisonic software tools like SPARTA and the IEM Plugin Suite. An example MATLAB script that implements this formulation is provided.
\end{abstract}

\section{Introduction}

This document is not intended to be self-sufficient. It is rather a supplement to~\citep{Ahrens:JASA2021}. We therefore recommend to the reader to familiarize with that document first and then turn to the present one. 

\section{Definitions}

The mathematical formulation of the equatorial microphone array (EMA) presented in~\citep{Ahrens:JASA2021} is strongly inspired by~\citep{Gumerov:book2005}, which causes some of the employed mathematical conventions to be incompatible with the most common formulation of ambisonics. More explicitly, we will derive ambisonic signals in this document that use the N3D normalization~\citep{Nachbar:AmbiX2011}, \citep[Sec.~4.7]{Zotter:book2019}. This requires the following three modifications to be applied to the original formulation of the EMA:
\begin{itemize}
    \item The definition of the spherical harmonics (SHs) needs to be a specific real one instead of a complex one
    \item The definition of the radial terms need to include the factor $4\pi\i^{n}$. See~\citep[Sec.~5.5]{Ahrens:binaural_rendering_in_sh} for a motivation.
\end{itemize}
The example MATLAB implementation~\citep{Ahrens:matlab_encoding} that accompanies this document uses the built-in function \texttt{fft.m} to compute the Fourier transform, which uses a negative exponent in the forward transform and a positive one in the inverse transform. This makes $h_n^{(2)}\!\left( \cdot \right)$, the spherical Hankel function of second kind, represent outward radiating sound fields. We refer the reader to~\citep{Ahrens:sign_conventions} for a more detailed discussion of this matter. And we refer the reader to~\citep{Ahrens:binaural_rendering_in_sh} for a more detailed discussion on conventions and definitions in the context of SH-based representations of sound fields.\\[1ex]

We define the orthonormal, real-valued circular harmonics (CH) $C_{m}(\alpha)$ as
\begin{equation}\label{eq:ch_definition}
    C_{m}(\alpha) = \begin{cases}
 \sqrt{2} \sin{(|m| \alpha)}, \ \forall m < 0\\
 \ \ \ \  \ \ \ 1 \ \ \ \ \ \ \ \ \, , \ \forall m = 0\\
 \sqrt{2} \cos{(m \alpha)}\ , \ \forall m > 0
\end{cases} \ . 
\end{equation}
Consequently, we define the orthonormal SHs $Y_{n,m}(\beta, \alpha)$ as~\cite[Sec.~18.4]{Whittaker:book1963}
\begin{equation}\label{eq:sh_definition}
    Y_{n,m}(\beta, \alpha) =
N_{n,m}(\beta) \, C_{m}(\alpha) \ ,
\end{equation}
with
\begin{equation}\label{eq:N_nm}
    N_{n,m}(\beta)= (-1)^m\, \sqrt{\frac{2n{+}1}{4\pi} \frac{(n{-}|m|)!}{(n{+}|m|)!}}\,P_n^{|{m}|}(\cos\beta)  \ .
\end{equation}
$\beta$ and $\alpha$ are the colatitude and azimuth angles of a spherical coordinate system, respectively. This definition of $Y_{n,m}(\beta, \alpha)$\footnote{Associated Legendre functions are used with different normalizations. We assume that $P_{n,m}(\mu)$ is defined via the following Rodriguez formula~\citep[Eq.~(2.1.20)-(2.1.21)]{Gumerov:book2005}:
\begin{equation}\nonumber
    P_{n,m}(\mu) = (-1)^m(1-\mu^2)^{m/2} \frac{\d^m}{\d \mu^m} P_n(\mu),  \ \ \forall n\geq 0, \ m\geq m \ ,
\end{equation}
with
\begin{equation}\nonumber
    P_n(\mu) = \frac{1}{2^n n!} \frac{\d^n}{\d \mu^n} (\mu^2 - 1),  \ \ \forall n\geq 0 \ .
\end{equation}
This also the definition that MATLAB and SciPy use.
} corresponds to the most commonly employed real-valued one and is used in the N3D ambisonic format~\citep{Nachbar:AmbiX2011}. It is also identical to the definition that is used in Politis's \texttt{getSH.m} MATLAB function~\citep{Politis:SH_transform}.

The sound pressure~$S^\text{\;\!int}(\beta, \alpha, r, \omega)$ in an interior domain i.e.,~the sound pressure inside a spherical domain centered at the coordinate origin that is free of sound sources and free of reflecting boundaries, can be represented in SHs in different ways~\citep{Gumerov:book2005}. To be consistent with the ambisonics formulation, we represent it as~\citep[Sec.~5.5]{Ahrens:binaural_rendering_in_sh}
\begin{eqnarray}\label{eq:sh_decomp_general}
    S^\text{\;\!int}(\beta, \alpha, r, \omega) &=& 
    \sum_{n=0}^\infty \sum_{m=-n}^n\!\! 4\pi\i^{n} \, \breve{S}_{n,m}(\omega)\,\,j_n\!\left( \omega \frac{r}{c} \right)\,Y_{n,m}(\beta, \alpha) \\
    &=& 
    \sum_{n=0}^\infty \sum_{m=-n}^n\!\! 4\pi\i^{n} \, \breve{S}_{n,m}(\omega)\,\,j_n\!\left( \omega \frac{r}{c} \right)\,N_{n,m}(\beta) \, C_{m}(\alpha) \ .
\end{eqnarray}
$\breve{S}_{n,m}(\omega)$ is the frequency-domain representation of the ambisonic signals $\breve{s}_{n,m}(t)$ that we would like to extract from the microphone signals.

The sound pressure $S(\beta, \alpha, R, \omega)$ on the equator of a sphere of radius $R$ is consequently given by
\begin{equation}\label{eq:sh_decomp_general_scat}
    S(\pi/2, \alpha, R, \omega) =
    \sum_{n=0}^\infty \sum_{m=-n}^n\!\! \breve{S}_{n,m}(\omega)\,\,b_n\!\left( \omega \frac{r}{c}, R \right)\, N_{n,m}(\pi/2) \, C_{m}(\alpha)
\end{equation}
with radial term 
\begin{equation}\label{eq:radial_filter_3d_wronskian}
    b_n\!\left( \omega \frac{R}{c}, R \right) = - 4\pi\i^n \, 
    \frac{\i}{\left(\omega\frac{R}{c}\right)^2} \frac{1}{h_n^{\prime\;\!(2)}\!\left( \omega \frac{R}{c} \right)} \ .
\end{equation}
$h_n^{\prime\;\!(2)}\!\left( \cdot \right)$ is the derivative of the spherical Hankel function $h_n^{(2)}\!\left( \cdot \right)$ with respect to the argument.

\section{Adaptation of the Derivation of the Solution}

When adapted to the CH definition~\eqref{eq:ch_definition}, \citep[Eq.~(10)]{Ahrens:JASA2021} reads:
\begin{equation}\label{eq:mathring_S_scat}
     \mathring{S}_m^\text{\;\!surf}(\pi/2, R, \omega) =
     \frac{1}{2\pi} \int_0^{2\pi}\!\! S^\text{\;\!surf}(\pi/2, \alpha, R, \omega)\,\, C_{m}(\alpha)\,\,\d\alpha \ .
\end{equation}   
\citep[Eq.~(12)]{Ahrens:JASA2021} reads then:
\begin{equation}\label{eq:sh_decomp_general_m}
    S^\text{\;\!surf}(\pi/2, \alpha, R, \omega) =
    \sum_{\phantom{|}m=-\infty}^\infty \sum_{n=|m|}^\infty\! \breve{S}_{n,m}(\omega)\,\, b_n\!\left( \omega \frac{R}{c}, R \right)\, N_{n,m}(\beta) \, C_{m}(\alpha) \ ,
\end{equation}
\citep[Eq.~(13)]{Ahrens:JASA2021} reads then:
\begin{equation}\label{eq:13}
    \mathring{S}_m^\text{\;\!surf}(R, \omega) = \!\!\sum_{n=|m|}^\infty\! \breve{S}_{n,m}(\omega)\,\, b_n\!\left( \omega \frac{R}{c}, R \right)\, N_{n,m}(\beta) \ .
\end{equation}
\citep[Eq.~(14)]{Ahrens:JASA2021} reads then:
\begin{equation}\label{eq:S_nm_pw}
   \breve{S}_{n,m}(\omega) = Y_{n,m}(\pi/2, \theta) = N_{n,m}(\pi/2) \, C_{m}(\theta) \ ,
\end{equation}
We introduce~\eqref{eq:S_nm_pw} into~\eqref{eq:13} to obtain the equivalent of~\citep[Eq.~(15)]{Ahrens:JASA2021}:
\begin{equation}\label{eq:s_ring_sh_pw}
   \mathring{S}_m^\text{\;\!surf}(R, \omega) = 
    C_{m}(\theta)\!\! \sum_{n=|m|}^\infty\!\! b_n\!\left( \omega \frac{R}{c}, R \right) \left[N_{n,m}(\pi/2)\right]^2 .
\end{equation}
The equivalent of~\citep[Eq.~(18)]{Ahrens:JASA2021} reads then:
\begin{equation}\label{eq:S_n_general_solution}
    \mathring{\bar{S}}_m(\omega) = 
    \frac{1}{\sum_{n'=|m|}^\infty \, b_{n'}\!\left( \omega \frac{R}{c}, R \right)\, \left[N_{n',m}(\pi/2)\right]^2} \ \mathring{S}_{m}^\text{\;\!surf}(R, \omega) \ .
\end{equation}
Complementing the equivalent of~\citep[Eq.~(17)]{Ahrens:JASA2021} given by
\begin{equation}\label{eq:sh_decomp_general_n}
    \mathring{S}_m^\text{\;\!surf}(R, \omega) =
    \sum_{n=|m|}^\infty\!\! \mathring{\bar{S}}_m(\omega) N_{n,m}(\pi/2) \, b_n\!\left( \omega \frac{R}{c}, R \right) N_{n,m}(\pi/2) \ .
\end{equation}
with~\eqref{eq:sh_decomp_general_scat} shows that our sought-after ambisonic signals $\breve{S}_{n,m}(\omega)$ are finally given by
\begin{equation}\label{eq:sh_decomp_general_n_2}
   \breve{S}_{n,m}(\omega) = \mathring{\bar{S}}_m(\omega) N_{n,m}(\pi/2) \ .
\end{equation}
\vspace{2ex}

The MATLAB scripts provided in~\citep{Ahrens:matlab_encoding} implement the above to compute the time-domain ambisonic signal from the microphone signals via the following steps:
\begin{itemize}
    \item Apply~\eqref{eq:mathring_S_scat} in the time domain
    \item Evaluate~\eqref{eq:S_n_general_solution} in time domain (whereby the multiplication between the two factors on the right-hand side turns into a convolution); the filter that is given by the first factor on the right-hand side of~\eqref{eq:S_n_general_solution} is computed in frequency domain and then transformed to time domain by means of an inverse Fourier transform and a modeling delay
    \item Evaluate~\eqref{eq:sh_decomp_general_n_2}
\end{itemize}
The result are real-valued ambisonic signals that can, for example, be rendered binaurally using standard tools such as SPARTA\footnote{\url{https://leomccormack.github.io/sparta-site/}} and the IEM Plugin Suite\footnote{\url{https://plugins.iem.at/}}. For this, the channels of the time-domain ambisonic signals $\breve{s}_{n,m}(t)$ from~\eqref{eq:sh_decomp_general_n_2} need to be ordered according to the \emph{Ambisonic Channel Number} (ACN). The channel number is given by $n^2+n+m$. Also make sure that the normalization is set to N3D in the software tool in which you are using the computed ambisonic signals.

\section{Rendering}

For completeness, we complement the processing pipeline here and illustrate how binaural rendering of the ambisonic signals $\breve{S}_{n,m}(\omega)$ and $\breve{s}_{n,m}(t)$, respectively, is performed, i.e., how the left and right ear signals $B^\text{L,R}(\omega)$ of a virtual listener in the captured sound field are computed. 

Taking into account the conventions used in the present report, the adapted Eq.~(9)\footnote{Note that Eq.~(9) from~\citep{Ahrens:JASA2021} comprises a mistake in that it lacks a factor of $\frac{1}{4\pi \i^{-n}}$ (cf.~\citep[Eq.~(23)]{Ahrens:binaural_rendering_in_sh}).} from~\citep{Ahrens:JASA2021} reads:
\begin{equation}\label{eq:binaural_rendering}
    B^\text{L,R}(\omega) =\sum_{n=0}^N \sum_{m=-n}^n\!\breve{S}_{n,m}(\omega)\,\,\mathring{H}_{n,m}^\text{L,R}(\omega) \ .
\end{equation}
$\mathring{H}_{n,m}^\text{L,R}(\omega)$ are the SH coefficients of the left and right far-field head-related transfer function $H^\text{L,R}(\beta, \alpha, \omega)$, respectively, defined as
\begin{equation}
    H^\text{L,R}(\beta, \alpha, \omega) =
    \sum_{n=0}^\infty \sum_{m=-n}^n\!\! \mathring{H}_{n,m}^\text{L,R}(\omega)\,Y_n^m(\beta, \alpha)
\end{equation}
$\beta$ and $\alpha$ are the colatitude and azimuth of the sound incidence direction. The MATLAB scripts provided in~\citep{Ahrens:matlab_encoding} also demonstrate how to implement~\eqref{eq:binaural_rendering}.



\end{document}